\documentclass[aps,prd,twocolumn,showpacs,nofootinbib]{revtex4-2}

\usepackage{graphicx}
\usepackage{amsmath}
\usepackage{amssymb}
\usepackage{bm}
\usepackage{bbm}
\usepackage{color}
\usepackage{slashed}
\usepackage[colorlinks=true,linkcolor=blue,citecolor=blue, urlcolor=blue]{hyperref}

\begin{document}

\title{Improved analysis of double $J/\psi$ production in $Z$-boson decay}

\author{Guang-Yu Wang$^{a}$}
\email{wanggy@cqu.edu.cn}
\author{Xing-Gang Wu$^{a}$}
\email{wuxg@cqu.edu.cn}
\author{Xu-Chang Zheng$^{a}$}
\email{zhengxc@cqu.edu.cn}
\author{Jiang Yan$^{a}$}
\email{yjiang@cqu.edu.cn}
\author{Jia-Wei Zhang$^{b}$}
\email{jwzhang@cqust.edu.cn}

\affiliation{$^a$ Department of Physics, Chongqing Key Laboratory for Strongly Coupled Physics, Chongqing University, Chongqing 401331, People's Republic of China \\
$^b$ Department of Physics, Chongqing University of Science and Technology, Chongqing, 401331, People's Republic of China}

\begin{abstract}
In this paper, we present an improved calculation for the decay rate of the rare $Z$-boson decay into $J/\psi + J/\psi$. This decay is dominated by the photon fragmentation mechanism, i.e., the transition $Z\to J/\psi + \gamma^{*}$ followed by the fragmentation $\gamma^{*}\to J/\psi$. In our calculation, the amplitude of $\gamma^{*}\to J/\psi$ is extracted from the measured value of $\Gamma(J/\psi \to e^+ e^-)$, and the amplitude of $Z\to J/\psi + \gamma^{*}$ is calculate through the light-cone approach. The higher-order QCD and relativistic corrections in the amplitude of $\gamma^{*}\to J/\psi$ and the large logarithms of $m_{_Z}^2/m_c^2$ that appear in the amplitude of $Z\to J/\psi + \gamma^{*}$ are resummed in our calculation. Besides, the non-fragmentation amplitude is calculated based on the NRQCD factorization, and the next-to-leading order QCD and relativistic corrections are included. The obtained branching fraction for this $Z$ decay channel is $8.66 ^{+1.48} _{-0.69}\times 10^{-11}$.
\end{abstract}

\maketitle

\section{Introduction}
\label{secIntro}

The rare $Z$-boson decay into double $J/\psi$ provides a good platform for studying the heavy quarkonium production and for searching of new physics beyond the standard model (SM). In 2019, the CMS collaboration presented the first search for the decay $Z \to J/\psi + J/\psi$ based on the data from $pp$ collisions with an integrated luminosity of $37.5\, {\rm fb}^{-1}$ at $\sqrt{s}=13\,{\rm TeV}$ at the large hadron collider (LHC). The observed upper limit on the branching fraction for the decay is ${\rm Br}(Z \to J/\psi + J/\psi)<2.2 \times 10^{-6}$ \cite{CMS:2019wch}. In 2022, the CMS collaboration updated the upper limit on this branching fraction by using a data set of $pp$ collisions corresponding to an integrated luminosity of $138\,{\rm fb}^{-1}$. Their updated upper limit is ${\rm Br}(Z \to J/\psi + J/\psi)<1.4 \times 10^{-6}$ \cite{CMS:2022fsq}. Although this decay process has not been observed up to now, the prospect of observing this decay at future colliders is optimistic \cite{dEnterria:2023wjq}. As is well-known, there are several lepton colliders, e.g., ILC \cite{ILC:2013jhg}, CEPC \cite{CEPCStudyGroup:2018ghi}, FCC-ee \cite{FCC:2018evy}, and super Z factory \cite{zfactory}, are under consideration. These lepton colliders are planed to operate at the $Z$ pole for a period of time, and a large number of $Z$ bosons are expected to be produced, e.g., at the FCC-ee, about $5\times 10^{12}$ $Z$ bosons will be produced \cite{FCC:2018evy}. Therefore, these lepton colliders provide new opportunities for studying the rare decay $Z \to J/\psi + J/\psi$.

On the theoretical side, the decay $Z \to J/\psi + J/\psi$ was first studied in Ref.\cite{Bergstrom:1990bu} in 1990. In Ref.\cite{Likhoded:2017jmx}, this decay was reanalysed at leading order (LO) in $\alpha_s$ and $v$ ($v$ stands for the typical velocity of the $c$ quark or the $\bar{c}$ quark in the rest frame of $J/\psi$) based on the nonrelativistic QCD (NRQCD) factorization. In these references, only the contribution of the LO QCD Feynman diagrams were considered. In Ref.\cite{Gao:2022mwa}, Gao and Gong pointed out that besides the LO QCD diagrams, the other transition via $Z \to J/\psi+\gamma^*$ followed by the fragmentation $\gamma^* \to J/\psi$ can also bring significant contribution. Their calculation shows that the branching fraction of $Z \to J/\psi + J/\psi$ can be enhanced to $1.1 \times 10^{-10}$ from $1.5 \times 10^{-13}$ after including the single photon fragmentation contribution. Recently, the next-to-leading order (NLO) QCD correction to the decay rate of $Z \to J/\psi + J/\psi$ has been calculated in Refs.\cite{Li:2023tzx,Luo:2022ugd}. The results there show that the NLO QCD correction can diminish the decay rate significantly \cite{Li:2023tzx}.

In this paper, we devote ourselves to presenting an improved calculation for the decay width of $Z \to J/\psi + J/\psi$. As the photon fragmentation contribution dominates this decay process, we first focus on the amplitude for the photon fragmentation diagrams. We express the amplitude of the photon fragmentation diagrams into two parts: one for $Z \to J/\psi+\gamma^*$, and the other for $\gamma^* \to  J/\psi$. For the amplitude of $\gamma^* \to  J/\psi$, we directly extract it from the experimentally measured leptonic decay width of the $J/\psi$, rather than calculating it order by order in $\alpha_s$ and $v$. Using this approach, the QCD and relativistic corrections to the amplitude of $\gamma^* \to  J/\psi$ are resummed up to all orders. For the amplitude of $Z \to J/\psi+\gamma^*$, we adopt the light-cone approach suggested by Refs.\cite{Lepage:1980fj, Chernyak:1983ej}, in which the amplitude is expressed as the convolution of the hard-scattering kernel with the $J/\psi$ light-cone distribution amplitude (LCDA). The hard-scattering kernel can be calculated with the perturbative QCD (pQCD), and the LCDA for $J/\psi$ can be calculated through the NRQCD factorization \cite{Bodwin:1994jh}. Under the light-cone approach, logarithms of $m_{_Z}^2/m_c^2$ appearing in the amplitude can be resummed through solving the evolution equation of the LCDA. In addition to the contribution of the photon fragmentation diagrams, we also calculate the contribution from the non-fragmentation diagrams. We shall adopt the fixed-order approach under the NRQCD factorization, which is adopted in previous calculations on the decay rate of $Z \to J/\psi + J/\psi$, to calculate the non-fragmentation amplitude.

The rest of the paper is organized as follows. In Sec.\ref{sec2}, we give the calculation formalism for the photon fragmentation amplitude, the non-fragmentation amplitude, and the decay width of $Z \to J/\psi + J/\psi$. In Sec.\ref{sec3}, we present the numerical results and discussions. Section \ref{sec4} is reserved as a summary.

\section{Calculation formalism}
\label{sec2}

\subsection{Fragmentation amplitude}

\begin{figure}[htb]
\includegraphics[width=0.45\textwidth]{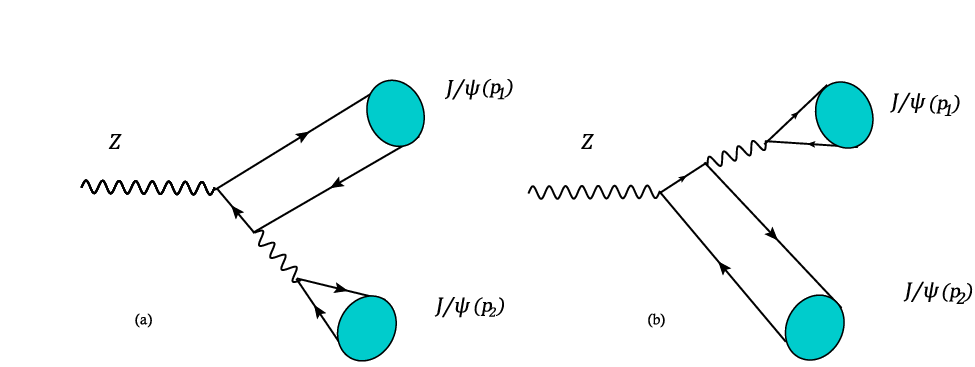}
\caption{Two typical LO Feynman diagrams for the photon fragmentation mechanism of the decay $Z \to J/\psi + J/\psi$.}
\label{fig.frag}
\end{figure}

The dominant mechanism of the decay $Z \to J/\psi + J/\psi$ is via the photon fragmentation, i.e., the transition $Z \to J/\psi+\gamma^*$ followed by $\gamma^* \to  J/\psi$. To understand this mechanism, we present two typical LO Feynman diagrams for the photon fragmentation mechanism in Fig.\ref{fig.frag} \footnote{It is noted that we shall adopt an improved approach where some important higher-order corrections are resummed, rather than the conventional fixed-order approach. Hence, Fig.\ref{fig.frag} is only used to illustrate the photon fragmentation mechanism, we do not use it in our calculation.}. The amplitude for the photon fragmentation mechanism can be written as
\begin{eqnarray}
&&i\mathcal{M}^{\rm fr}_{Z\to J/\psi+ J/\psi}\nonumber \\
&&=i\mathcal{M}^{\mu}_{Z\to J / \psi(P_1)+\gamma^{*}(P_2 )}\frac{-i}{m_{J/\psi}^2} i{\cal M}_{\gamma^* \to J/\psi,\mu} \nonumber \\
&&~~~+(1\leftrightarrow 2),
\label{eq.ampfrag}
\end{eqnarray}
where the factor $-i/m_{J/\psi}^2$ arises from the photon propagator that will fragment into a $J/\psi$. $i{\cal M}_{\gamma^* \to J/\psi,\mu}$ denotes the amplitude for the transition of a virtual photon into the $J/\psi$, and it can be expressed as \cite{Bodwin:2006yd,Bodwin:2013gca}
\begin{eqnarray}
&& i{\cal M}^{\mu}_{\gamma^* \to J/\psi}\nonumber \\
&& =-ie\langle J/\psi|J^{\mu}(x=0)|0\rangle=-i e g_{J/\psi\gamma}\epsilon_{J/\psi}^{*\mu},
\label{eq.amp-gamma-jpsi}
\end{eqnarray}
where $J^{\mu}(x)$ is the electromagnetic current,
\begin{eqnarray}
J^{\mu}(x)= \sum_{q} e_{q} \bar{q}(x)\gamma^{\mu}q(x).
\label{eq.EMcurrent}
\end{eqnarray}		
Here, the sum extends over all quark flavors. The $J/\psi$-photon effective coupling constant $g_{J/\psi\gamma}$ can be determined using the decay width of the $J/\psi$ to an electron-positron pair \cite{Bodwin:2006yd}, i.e.,
\begin{eqnarray}
\Gamma(J/\psi \to e^{+}e^{-})=\frac{4\pi\alpha^2(m_{J/\psi})g_{J/\psi\gamma}^2}{3m_{J/\psi}^3}.
\label{eq.width-jpsi-lepton}
\end{eqnarray}
The method of calculating the amplitude $i{\cal M}_{\gamma^* \to J/\psi,\mu}$ based on Eqs.(\ref{eq.amp-gamma-jpsi}), (\ref{eq.EMcurrent}) and (\ref{eq.width-jpsi-lepton}) has an important advantage compared to the fixed-order calculation based on the NRQCD factorization. Under the NRQCD factorization, an amplitude is expressed as a double expansion in $\alpha_s$ and $v$. The calculations of higher-order QCD and relativistic corrections are very difficult. On the other hand, the calculation of the amplitude $i{\cal M}_{\gamma^* \to J/\psi,\mu}$ based on Eqs.(\ref{eq.amp-gamma-jpsi}), (\ref{eq.EMcurrent}) and (\ref{eq.width-jpsi-lepton}) already contains the QCD and relativistic corrections to all orders.

Next, let us demonstrate the calculation of the amplitude $\mathcal{M}^{\mu}_{Z\to J / \psi+\gamma^{*}}$. We shall adopt the approximation $\mathcal{M}^{\mu}_{Z\to J / \psi+\gamma^{*}}\approx \mathcal{M}^{\mu}_{Z\to J / \psi+\gamma}$, which will only lead to an error of ${\cal O}(m_{J/\psi}^2/m_{_Z}^2)$. The decay $Z\to J/\psi +\gamma$ has been studied extensively based on the NRQCD and light-cone approaches \cite{Guberina:1980dc,Luchinsky:2017jab,Wang:2013ywc,Bodwin:2017pzj,Sang:2023hjl,Wang:2023ssg}. In this work, we adopt the light-cone approach to calculate the amplitude $\mathcal{M}^{\mu}_{Z\to J / \psi+\gamma}$.

Under the light-cone approach, the amplitude of $Z\to J / \psi+\gamma$ is expressed by \cite{Bodwin:2017pzj}
\begin{eqnarray}
i\mathcal{M}_{Z\to J/\psi+\gamma,\mu}=i\mathcal{A}\, \epsilon_{\xi\mu\nu\rho}\epsilon_{Z}^{\xi}
\epsilon_{J/\psi}^{*\nu}p_{\gamma}^{\rho},
\end{eqnarray}
where
\begin{eqnarray}
i\mathcal{A}=-\frac{e e_c g_{_Z} g_A^c m_{J/\psi}}{m_{_Z}^2}f_{J/\psi}^{\parallel} \int_{0}^1 dx\, T_H(x,\mu)\phi_{J/\psi}^{\parallel}(x,\mu).
\label{eq.convol}
\end{eqnarray}
Here, $e_c=+2/3$ is the electric charge of the charm quark, $f_{J/\psi}^\parallel$ is the decay constant of the longitudinally polarized $J/\psi$, $T_H(x,\mu)$ is the hard-scattering kernel, $\phi_{J/\psi}^{\parallel}$ is the longitudinally polarized LCDA of the $J/\psi$, $\epsilon_Z$ is the polarization vector of the $Z$ boson, $\epsilon^{*}_{J/\psi}$ is the polarization vector of $J/\psi$, and $p_\gamma$ is the momentum of the photon, $\mu$ is the factorization scale, and $x$ is the momentum fraction of the $c$ quark in the $J/\psi$. The couplings $g_{_Z}$ and $g_A^c$ are defined as $g_{_Z} = 2(\sqrt{2}G_F)^{1/2}m_{_Z}$, $g_A^c = \frac{1}{2}t_{3L}^c$, where $G_F$ is the Fermi constant and $t_{3L}^c=1/2$ is the weak isospin for the left-handed $c$ quark.

The hard-scattering kernel $T_H(x,\mu)$ can be calculated through pQCD, and it has been calculated up to order $\alpha_s$ in Ref.\cite{Wang:2013ywc}. The expression of the hard-scattering kernel, up to order $\alpha_s$, is
\begin{equation}
T_H(x,\mu)=
T_H^{(0)}(x,\mu)+\frac{\alpha_s(\mu)}{4\pi} T_H^{(1)}(x,\mu),
\end{equation}
where
\begin{align}\label{eq.TH-x}
T_H^{(0)}(x,\mu) =&\, \frac{1}{x(1-x)},\notag\\
T_H^{(1)}(x,\mu) =&\, C_F\frac{1}{x(1-x)} \bigg\{ \big[3+2x\,{\rm ln}(1-x)\notag\\
&\,+2(1-x)\,{\rm ln} x\big]\left({\rm ln}\frac{m_{_Z}^2}{\mu^2}-i\pi\right)\notag\\
&\,+x\,{\rm ln}^2(1-x)+(1-x)\,{\rm ln}^2 x\notag\\
&\,-(1-x)\,{\rm ln}(1-x)-x\,{\rm ln} x-9
\bigg\}.
\end{align}
Here, $C_F=(N_c^2-1)/(2N_c)$ with $N_c=3$ is the quadratic Casimir operator. In order to avoid large logarithms appearing in the hard-scattering kernel, we set the factorization scale as $\mu=m_{_Z}$ in Eq.(\ref{eq.convol}).

The longitudinally polarized LCDA $\phi_{J/\psi}^\parallel$ is defined as \cite{Ball:1996tb,Ball:1998sk}:
\begin{align}
&\langle J/\psi(p)|\bar{Q}(z)\gamma^\mu [z,0] Q(0) |0 \rangle
\nonumber \\
=&-p^\mu \frac{\epsilon_{J/\psi}^*\cdot z}{p\cdot z}f_{J/\psi}^\parallel m_{J/\psi}\int_0^1 dx\, e^{ip\cdot zx}\phi_{J/\psi}^\parallel(x,\mu),
\label{eq.LCDA-defn}
\end{align}
where $p$ is the quarkonium momentum, the LCDA is normalized as $\int_0^1 dx\,\phi_{J/\psi}^\parallel(x,\mu)=1$. The coordinate $z$ lies along the plus light-cone direction, and the gauge link $[z,0]$ that makes the definition of the LCDA gauge invariant is
\begin{eqnarray}
[z,0]=P\, {\rm exp}\left[i g_s\int_0^z dx A_a^+(x) T^a  \right],
\end{eqnarray}
where $P$ denotes the path ordering.

Under the NRQCD factorization, the LCDA $\phi_{J/\psi}^\parallel$ is expressed as a double expansion in $\alpha_s$ and $v$ \cite{Wang:2013ywc, Ma:2006hc}, i.e.,
\begin{eqnarray}
\phi_{J/\psi}^\parallel(x,\mu_0)&&= \phi_{J/\psi}^{\parallel(0)}(x,\mu_0)+\langle v^2\rangle_{J/\psi}\phi_{J/\psi}^{\parallel(v^{2})}(x,\mu_0)	\nonumber \\
&& +\frac{\alpha_s(\mu_0)}{4\pi} \phi_{J/\psi}^{\parallel(1)}(x,\mu_0)+{\cal O}(\alpha_s^2,\alpha_s v^2,v^4).\nonumber\\
\end{eqnarray}
Here, in order to avoid large logarithms appearing in the LCDA, we set the initial factorization scale of the LCDA as $\mu_0=m_c$. The quantity $\langle v^2\rangle_{J/\psi}$ is a ratio of NRQCD LDMEs, whose definition is
\begin{equation}
\langle v^{2}\rangle_{J/\psi}=\frac{\langle \bm{q}^{2}\rangle_{J/\psi}}{m_c^2}=\frac{\langle J/\psi(\lambda)|\psi^\dagger(-\frac{i}{2}
\tensor{\bm{D}})^{2}\bm{\sigma}\cdot\bm{\epsilon}(\lambda)\chi|0\rangle}{m_c^{2}\,\langle J/\psi(\lambda)
|\psi^\dagger \bm{\sigma}\cdot\bm{\epsilon}(\lambda)\chi|0\rangle},
\label{eq.v2-def}
\end{equation}
where the covariant derivative operator $\psi^\dagger \tensor{\bf D}\chi\equiv \psi^\dagger {\bf D}\chi-({\bf D} \psi)^\dagger \chi$, $\psi$ and $\chi$ are Pauli spinor fields that describe $c$ quark annihilation and $\bar{c}$ quark creation, respectively.

The expression for the LO LCDA is \cite{Wang:2013ywc}
\begin{eqnarray}
\phi_{J/\psi}^{\parallel(0)}(x,\mu_0) =&\,	\delta(x-\tfrac{1}{2}).
\end{eqnarray}
The order-$v^2$ LCDA $\phi_{J/\psi}^{\parallel(v^{2})}(x,\mu_0)$ was calculated in Ref.\cite{Wang:2017bgv}, and the order-$\alpha_s$ LCDA $\phi_{J/\psi}^{\parallel(1)}(x,\mu_0)$ was calculated in Ref.\cite{Wang:2013ywc}. Their expressions are
\begin{align}
\phi_{J/\psi}^{\parallel(v^2)}(x,\mu_0) =&\, \frac{\delta^{(2)}(x-\frac{1}{2})}{24}, \label{phi-vsq}\\
\phi_{J/\psi}^{\parallel(1)}(x,\mu_0) =&\,	C_F\theta(1-2x) \Bigg\{ \Bigg[ \left( 4x + \frac{8x}{1-2x} \right)\notag\\
&\, \cdot \left(\log\frac{\mu_0^2}{m_c^2(1-2x)^2}-1\right)\Bigg]_{+} - \left[8x\right]_{+} \notag\\
&\, +\left[\frac{16x(1-x)}{(1-2x)^2}\right]_{++}\Bigg\} + (x\leftrightarrow 1-x), \label{eq:phiV-alphas}
\end{align}
where $\delta^{(2)}$ stands for the second derivative of the delta function. The $+$ and $++$ functions are defined as
\begin{align}\label{def:+and++functions}
\int_0^1 {\rm d}x\,[f(x)]_{+}g(x)=&\,\int_0^1 {\rm d}x\, f(x)\left[g(x)-g(1/2)\right], \\
\int_0^1 {\rm d}x\,[f(x)]_{++}g(x)=&\,\int_0^1 {\rm d}x\, f(x) [g(x) - g(1/2) \notag\\
&\,-g'(1/2)(x-1/2)].
\end{align}

To calculate the amplitude based on Eq.(\ref{eq.convol}), we need the LCDA at the scale $\mu=m_{_Z}$. We calculate the LCDA $\phi_{J/\psi}^{\parallel}(x,\mu=m_{_Z})$ through solving the evolution equation of the LCDA, where the initial LCDA $\phi_{J/\psi}^{\parallel}(x,\mu_0=m_c)$ is used as the boundary condition. The scale evolution of the LCDA $\phi_{J/\psi}^\parallel(x,\mu)$ obeys the Efremov-Radyushkin-Brodsky-Lepage (ERBL) equation \cite{Lepage:1980fj,Efremov:1979qk,Efremov:1978rn}, i.e.,
\begin{eqnarray}
\frac{\partial}{\partial \,{\rm ln}\, \mu^2}\phi_{J/\psi}^\parallel(x,\mu)=\int_0^1 dy\,V_\parallel[x,y;\alpha_s(\mu)]\, \phi_{J/\psi}^\parallel(y,\mu).
\label{eq.ERBL}
\end{eqnarray}
where the evolution kernel can be expanded in $\alpha_s$ as
\begin{eqnarray}
&& V_\parallel[x,y;\alpha_s(\mu)]\nonumber \\
&& =\frac{\alpha_s(\mu)}{4\pi}V^{(0)}(x,y)+\left(\frac{\alpha_s(\mu)}{4\pi}\right)^2 V^{(1)}(x,y)\nonumber \\
&& ~~~+{\cal O}(\alpha_s^2),
\end{eqnarray}
At present, the order-$\alpha_s$ evolution kernel $V^{(0)}(x,y)$ is given in Refs.\cite{Lepage:1980fj,Efremov:1979qk}, and the order-$\alpha_s^2$ evolution kernel $V^{(1)}(x,y)$ is given in Refs.\cite{Dittes:1983dy,Katz:1984gf,Mikhailov:1984ii}.

To solve the ERBL equation, i.e., Eq.(\ref{eq.ERBL}), it is convenient to expand the LCDA in terms of the Gegenbauer polynomials $C_n^{(3/2)}(2x-1)$, which are the eigenfunctions of the LO evolution kernel. The Gegenbauer expansion of the LCDA is
\begin{eqnarray}
\phi_{J/\psi}^\parallel(x,\mu)= \sum_{n=0}^\infty a_n^\parallel(\mu)\, x(1-x)\,C_n^{(3/2)}(2x-1),
\end{eqnarray}
where $a_n^\parallel(\mu)$ is the $n_{\rm th}$ Gegenbauer moment of the LCDA $\phi_{J/\psi}^\parallel(x,\mu)$, and
\begin{eqnarray}
a_n^\parallel(\mu)=N_n\int_0^1 dx\,C_n^{(3/2)}(2x-1) \phi_{J/\psi}^\parallel(x,\mu),
\end{eqnarray}
where the normalization factor
\begin{eqnarray}
N_n=\frac{4(2n+3)}{(n+1)(n+2)}.
\end{eqnarray}

Solving the ERBL equation in the Gegenbauer moment space, the obtained the evolved Gegenbauer moments have the following form:
\begin{equation}
a_n^\parallel(\mu)=\sum_{k=0}^n U_{nk}(\mu,\mu_0)a_k^\parallel(\mu_0).
\end{equation}
The expressions for $U_{nk}(\mu,\mu_0)$ up to next-leading-logarithmic (NLL) accuracy can be found in Ref.\cite{Agaev:2010aq}.

Similarly, the hard-scattering kernel can also be expanded in terms of the Gegenbauer polynomials as
\begin{eqnarray}
T_H(x,\mu)=\sum_{n=0}^\infty N_n b_n(\mu) C_n^{(3/2)}(2x-1).
\end{eqnarray}
The $n_{\rm th}$ Gegenbauer moment
\begin{eqnarray}
b_n(\mu)=\int_0^1 dx\, x(1-x)C_{n}^{(3/2)}(2x-1) T_H(x,\mu).
\end{eqnarray}

Making use of the Gegenbauer expansions of the LCDA and the hard-scattering kernel and the orthogonality property of the Gegenbauer polynomials, we can express the convolution of the hard-scattering kernel and the LCDA in Eq.(\ref{eq.convol}) as
\begin{eqnarray}
&&\int_0^1 dx\, T_H(x,\mu)\phi_{J/\psi}^{\parallel}(x,\mu)\nonumber\\
&&=\sum_{n=0}^{\infty} b_n(\mu) a_n(\mu) \nonumber \\
&&=\sum_{n=0}^{\infty}\sum_{k=0}^{n} b_n(\mu)  U_{nk}(\mu,\mu_0) a_k(\mu_0),
\label{eq.ampsum}
\end{eqnarray}
As stated below Eq.(\ref{eq.TH-x}), we have set $\mu=m_{_Z}$ so that the hard-scattering kernel $T_H(x,\mu)$ does not contain large logarithms. The large logarithms of $m_{_Z}^2/m_c^2$ will appear in the LCDA $\phi_{J/\psi}^{\parallel}(x,\mu=m_{_Z})$. In our calculation, these large logarithms are resummed up to NLL accuracy through the evolution of the LCDA from the initial scale $\mu_0=m_c$ to the scale $\mu=m_{_Z}$.

It is found that the sum over $n$ in Eq.(\ref{eq.ampsum}) is divergent \cite{Bodwin:2017pzj}. In order to address the nonconvergent problem in the Gegenbauer moment summation, the authors of Ref.\cite{Bodwin:2016edd} introduced the Abel-Pad\'{e} method, which combines the Abel summation and the Pad\'{e} approximant. Under this method, the summation in Eq.(\ref{eq.ampsum}) is first written as an Abel summation as
\begin{eqnarray}
&&\sum_{n=0}^{\infty}\sum_{k=0}^{n} b_n(\mu)  U_{nk}(\mu,\mu_0) a_k(\mu_0)\nonumber \\
&&=\lim_{z\to 1}\sum_{n=0}^{\infty}\sum_{k=0}^{n} b_n(\mu) z^n U_{nk}(\mu,\mu_0) a_k(\mu_0).
\label{eq.Abelsum}
\end{eqnarray}
Then, we keep 20 nonzero terms in the second line of Eq.(\ref{eq.Abelsum}), and generate a $10 \times 10$ Pad\'{e} approximant. It is worth noting that the limit of $z \to 1$ should be taken after the generation of the Pad\'{e} approximant.

In addition to calculating the convolution in Eq.(\ref{eq.convol}), we also need to determine the value of the decay constant $f_{J/\psi}^{\parallel}$. Setting $z$ to 0 in Eq.(\ref{eq.LCDA-defn}) and using the normalization $\int_0^1 dx\,\phi_{J/\psi}^\parallel(x,\mu)=1$, we obtain
\begin{equation}
\langle J/\psi|\bar{c}(0)\gamma^\mu c(0) |0\rangle=-f_{J/\psi}^\parallel m_{J/\psi} \epsilon_{J/\psi}^{*\mu}.
\label{eq.decay-constant-def}
\end{equation}
With this relation, we can relate the decay constant $f_{J/\psi}^\parallel$ with the leptonic decay width of $J/\psi$ \cite{Bodwin:2017pzj}, i.e.,
\begin{equation}
\Gamma(J/\psi\to e^{+}e^{-})=\frac{4\pi}{3m_{J/\psi}}\alpha^2(m_{J/\psi})e_c^2 f_{J/\psi}^{\parallel 2}.
\label{eq.EM-width-decay-constant}
\end{equation}
The decay constant $f_{J/\psi}^\parallel$ can be determined through this relation straightforwardly.

\subsection{Non-fragmentation amplitude}

\begin{figure}[htbp]
\includegraphics[width=0.45\textwidth]{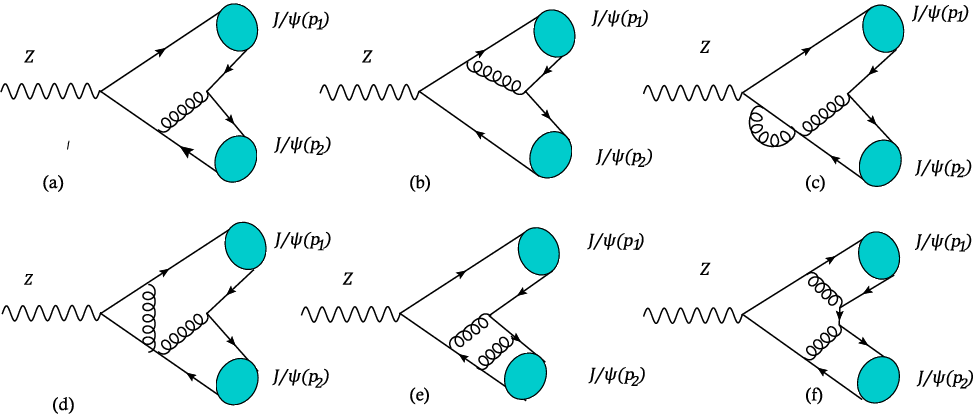}
\caption{Six typical tree-level and one-loop non-fragmentation Feynman diagrams that contribute to the decay $Z\to J/\psi +J/\psi$.}
\label{fig.nonfrag}
\end{figure}

In this subsection, we present the key formulas for calculating the non-fragmentation amplitude. Compared to the fragmentation amplitude, the non-fragmentation amplitude is significantly suppressed \cite{Gao:2022mwa,Li:2023tzx}. We will calculate it directly based on the NRQCD factorization.

Under the NRQCD factorization~\cite{Bodwin:1994jh}, the non-fragmentation amplitude up to order-$v^2$ for the decay $Z\to J/\psi +J/\psi$ can be written as
\begin{eqnarray}
&&{\cal M}^{\rm nfr}_{Z \to J/\psi +J/\psi}\nonumber \\
&&=2m_{J/\psi} \left(c_0+c_2 \langle v^2 \rangle_{J/\psi} \right)\langle J/\psi \vert \psi^\dagger {\bm \sigma}\cdot {\bm \epsilon} \chi \vert 0 \rangle^2,
\label{eq.nrqcdfact}
\end{eqnarray}
where $c_0$ and $c_2$ are short-distance coefficients (SDCs) that can be expanded in powers of $\alpha_s$. In this work, we calculate the SDC $c_0$ up to NLO in $\alpha_s$, and calculate the SDC $c_2$ at LO in $\alpha_s$. $\langle J/\psi \vert \psi^\dagger {\bm \sigma}\cdot {\bm \epsilon} \chi \vert 0 \rangle$ is the LDME for $J/\psi$. $\langle v^2 \rangle_{J/\psi}$ is a ratio of LDMEs, which has been defined in Eq.(\ref{eq.v2-def}). The factor $2m_{J/\psi}$ arises from that we use the relativistic normalization for the $J/\psi$ in ${\cal M}^{\rm nfr}_{Z \to J/\psi +J/\psi}$ on the left-hand side of Eq.(\ref{eq.nrqcdfact}), while we use conventional nonrelativistic normalization for the NRQCD matrix element on the right-hand side of Eq.(\ref{eq.nrqcdfact}).

The SDCs can be determined through the matching method. According to this method, we first replace the $J/\psi$ state in Eq.(\ref{eq.nrqcdfact}) by an on-shell $(c\bar{c})$ pair with a small relative velocity $v$, i.e.,
\begin{eqnarray}
&&{\cal M}^{\rm nfr}_{Z \to (c\bar{c})[^3S_1^{[1]}] +(c\bar{c})[^3S_1^{[1]}]}\nonumber \\
&&=\left(c_0+c_2 \,v^2 \right)\langle (c\bar{c})[^3S_1^{[1]}] \vert \psi^\dagger {\bm \sigma}\cdot {\bm \epsilon} \chi \vert 0 \rangle^2,
\label{eq.matching}
\end{eqnarray}
where the amplitude ${\cal M}^{\rm nfr}_{Z \to (c\bar{c})[^3S_1^{[1]}] +(c\bar{c})[^3S_1^{[1]}]}$ can be calculated using perturbative theory in QCD, and the LDME $\langle (c\bar{c})[^3S_1^{[1]}] \vert \psi^\dagger {\bm \sigma}\cdot {\bm \epsilon} \chi \vert 0 \rangle$ for an on-shell $(c\bar{c})$ pair can be calculated using perturbative theory in NRQCD. Then the SDCs can be determined by comparing both sides of Eq.(\ref{eq.matching}) order by order. The SDCs are insensitive to the long-distance physical effects, thus the SDCs extracted from the on-shell $(c\bar{c})$ pair production can be applied to the $J/\psi$ production.

To calculate the amplitude ${\cal M}^{\rm nfr}_{Z \to (c\bar{c})[^3S_1^{[1]}] +(c\bar{c})[^3S_1^{[1]}]}$ up to order $\alpha_s$ and order $v^2$, we assign the momenta of the $c$ and $\bar{c}$ quarks as $p_i=\frac{1}{2}P_i+q_i$ and $\bar{p}_i=\frac{1}{2}P_i-q_i$, where $i=1,2$. Then we have the relations
\begin{eqnarray}
&&p_i^2=\bar{p}_i^2=m_c^2,\, P_i^2=4 E_q^2,\, P_i\cdot q_i=0,\nonumber \\
&& q_i^2=m_c^2-E_q^2=-m_c^2 v^2.
\end{eqnarray}
In the rest frame of a $(c\bar{c})$ pair, the total and relative momenta of the $(c\bar{c})$ pair are $P_i=(2E_q,\bm{0})$ and $q_i=(0,\bm{q}_i)$ where $E_q=\sqrt{\bm{q}_i^2+m_c^2}$. In deriving these relations, we have implicitly adopted the relation $\bm{q}_1^2=\bm{q}_2^2=m_c^2 v^2$.

It is convenient to adopt the covariant projection method to enforce the produced $(c\bar{c})$ pairs to be the corresponding spin and color state. The spin projector for the spin-triplet state is \cite{Bodwin:2010fi}
\begin{eqnarray}
\Pi=&&\frac{1}{4\sqrt{2}E_q(E_q+m_c)}(\slashed{\bar{p}}_i-m_c)\slashed{\epsilon}^*(\slashed{P}_i+2E_q)\nonumber \\
&& \cdot (\slashed{p}_i+m_c),
\end{eqnarray}
where $\epsilon^*$ is the polarization vector of the spin-triplet state. The color projector for color-singlet state is
\begin{eqnarray}
\Lambda_1=\frac{\bf 1}{\sqrt{3}},
\end{eqnarray}
where ${\bf 1}$ stands for the unit matrix of the $SU(3)_c$ group.

To obtain the amplitude in terms of powers of $v$, we first expand the amplitude in powers of $q_i$ to the required order,
\begin{eqnarray}
&&{\cal M}^{\rm nfr}_{Z \to (c\bar{c})[^3S_1^{[1]}] +(c\bar{c})[^3S_1^{[1]}]}\nonumber \\
&&= {\cal M}\Big{|}_{q_1=q_2=0}+\sum_{i=1}^{2} q_i^{\alpha} \frac{\partial {\cal M}}{\partial q_i^{\alpha}}\Big{|}_{q_1=q_2=0}\nonumber \\
&&~~~ +\frac{1}{2!}\sum_{i,j=1}^{2}q_i^{\alpha}q_j^{\beta}\frac{\partial^{2}{\cal M}}{\partial q_i^{\alpha} \partial q_j^{\beta}}\Big{|}_{q_1=q_2=0} +\cdots,
\label{eq.expand1}
\end{eqnarray}
where ${\cal M}^{\rm nfr}_{Z \to (c\bar{c})[^3S_1^{[1]}] +(c\bar{c})[^3S_1^{[1]}]}$ is the non-fragmentation amplitude, which has been abbreviated as ${\cal M}$ on right hand side of Eq.(\ref{eq.expand1}). To project out the $S$-wave contribution, we make the replacement
\begin{eqnarray}
q_i^{\mu}q_i^{\nu} \to \frac{{\bm q}_i^2}{(D-1)}\Pi^{\mu \nu}.
\end{eqnarray}
where
\begin{eqnarray}
\Pi^{\mu \nu}=-g^{\mu \nu}+\frac{P_i^{\mu}P_i^{\nu}}{P_i^2},
\end{eqnarray}
The odd-power terms of $q_i$ vanish for the S-wave contribution. Then we obtain the amplitude up to order $v^2$,
\begin{eqnarray}
&&{\cal M}^{\rm nfr}_{Z \to (c\bar{c})[^3S_1^{[1]}] +(c\bar{c})[^3S_1^{[1]}]}\nonumber \\
&&= {\cal M}\Big{|}_{q_1=q_2=0}+\frac{1}{2!}\sum_{i=1}^{2}\frac{{\bm q}_i^2\Pi_{\alpha\beta}}{(D-1)}\frac{\partial^{2}{\cal M}}{\partial q_i^{\alpha} \partial q_i^{\beta}}\Big{|}_{q_1=q_2=0}.
\label{eq.expand2}
\end{eqnarray}

In the calculation, we expand the amplitude in powers of the relative momenta $q_i$ before carrying out the loop integration. This amounts to calculating the contributions arising from the hard region in the language of the strategy of region \cite{Beneke:1997zp}. Consequently, the Coulomb divergences arising from the potential region do not appear in our calculation.

There are ultraviolet (UV) and infrared (IR) divergences in the loop-diagram calculation. We utilize dimensional regularization to regularize both the UV and IR divergences, with spacetime dimension $D=4-2\epsilon$. The UV divergences are removed through renormalization. We define the renormalized charm-quark field, charm-quark mass, and gluon field in the on-shell (OS) scheme, while define the renormalized strong coupling constant in the modified minimal subtraction ($\overline{\rm MS}$) scheme. The corresponding renormalization constants are $Z_i=1+\delta Z_i$, and
\begin{eqnarray}
 \delta Z_2^{\rm OS} =&& -C_F\dfrac{\alpha_s}{4\pi}\left[\dfrac{1}{\epsilon_{UV}} +\frac{2}{\epsilon_{IR}} -3\gamma_E +3\ln\frac{4\pi \mu_r^2}{m_c^2}  +4 \right] , \nonumber\\
 \delta Z_m^{\rm OS} =&& -3C_F\dfrac{\alpha_s}{4\pi}\left[\dfrac{1}{\epsilon_{UV}} -\gamma_E +\ln\dfrac{4\pi \mu_r^2}{m_c^2} +\frac{4}{3}\right] ,\nonumber \\
 \delta Z_3^{\rm OS} =&& \frac{\alpha_s}{4\pi}\bigg[(\beta'_0-2C_A)\left(\dfrac{1}{\epsilon_{UV}} -\frac{1}{\epsilon_{IR}}\right)\nonumber  \\ &&-\frac{4}{3}T_F\left(\dfrac{1}{\epsilon_{UV}} -\gamma_E +\ln\dfrac{4\pi \mu_r^2}{m_c^2}\right) \bigg] , \nonumber\\
 \delta Z_g^{\overline{\mathrm{MS}}}= && -\dfrac{\beta_0}{2}\dfrac{\alpha_s}{4\pi}\left[\dfrac{1}{\epsilon_{UV}} -\gamma_E +\ln(4\pi) \right],
\end{eqnarray}
where $\gamma_E$ is the Euler constant, $\mu_r$ is the renormalization scale, $\beta_0=11/3-2 n_f/3$, $\beta'_0=11/3-2 n_{lf}/3$, $n_f$ is the number of active quark flavors and $n_{lf}=3$ is the number of light quark flavors.

In dimensional regularization, the $\gamma_5$ matrix should be noted. In this work, we employ the Larin scheme \cite{Larin:1993tq}, where the product of a $\gamma_{\mu}$ matrix and a $\gamma_5$ matrix is expreesed as:
\begin{eqnarray}
\gamma_\mu\gamma_5=\frac{i}{6}\epsilon_{\mu\rho\sigma\tau}\gamma^{\rho} \gamma^{\sigma} \gamma^{\tau}.
\end{eqnarray}
In this $\gamma_5$ scheme, a finite renormalization is required to restore the axial current Ward identity for the axial vector vertex. The corresponding finite renormalization constant $Z_5$ up to order $\alpha_s$ for the axial vector vertex is given by:
\begin{eqnarray}
Z_5=1-\frac{\alpha_s}{\pi}C_F.
\end{eqnarray}

In order to obtain the SDCs using Eq.(\ref{eq.matching}), in addition to the amplitude ${\cal M}^{\rm nfr}_{Z \to (c\bar{c})[^3S_1^{[1]}] +(c\bar{c})[^3S_1^{[1]}]}$, we also need to calculate the matrix element $\langle (c\bar{c})[^3S_1^{[1]}]  \vert \psi^\dagger {\bm \sigma}\cdot {\bm \epsilon} \chi \vert 0 \rangle$. The matrix element up to order $\alpha_s$ is
\begin{eqnarray}
\sum_{\lambda} \vert \langle (c\bar{c})[^3S_1] \vert \psi^\dagger{\bm \sigma}\cdot {\bm \epsilon}(\lambda) \chi \vert 0 \rangle \vert^2=2N_c(D-1)(2E_q)^2,
\end{eqnarray}
where $\lambda$ denotes the helicity of the $(c\bar{c})$ pair.

In the calculation of the non-framentation amplitude, the packages FeynArts \cite{feynarts} is adopted to generate Feynman diagrams, the package Feyncalc \cite{feyncalc1,feyncalc2} is adopted to carry out the traces over the Dirac and color matrices, the package \$Apart \cite{apart} is adopted to do partial fraction, and the package FIRE \cite{fire} is adopted to do integration-by-parts (IBP) reduction for the loop integrals. After the IBP reduction, the one-loop integral are reduced to mater integrals, and these master integrals are calculated numerically using the package LoopTools \cite{looptools}.

\subsection{Decay width}

The total amplitude for the decay $Z\to J/\psi +J/\psi$ is the sum of the fragmentation and non-fragmentation amplitudes, i.e.,
\begin{eqnarray}
\mathcal{M}_{Z\to J/\psi+ J/\psi} = &\, \mathcal{M}^{\rm fr}_{Z\to J/\psi+ J/\psi}+ \mathcal{M}^{\rm nfr}_{Z\to J/\psi+ J/\psi}.
\end{eqnarray}
Then the decay width can be calculated through
\begin{eqnarray}
&& \Gamma_{Z\to J/\psi+ J/\psi}\nonumber \\
&& =\frac{1}{2!}\frac{1}{3}\frac{\sqrt{m_{_Z}^2/4-m_{J/\psi}^2}}{8\pi m_{_Z}^{2}}\sum\left|\mathcal{M}_{Z\to J/\psi+ J/\psi}\right|^{2},
\end{eqnarray}
where the factor $1/2!$ comes from that there are two identical particles in the final state, $1/3$ comes from the average over the polarizations of the $Z$ boson, $\sum$ denotes that the polarizations of the initial and final states are summed over.

\section{Numerical results and discussions}
\label{sec3}

In the numerical calculation, the necessary input parameters are taken as follows:
\begin{eqnarray}
&& m_{_Z}=91.1876\,{\rm GeV}, \;\;\; m_c=1.4 \pm 0.2 \,{\rm GeV},\nonumber \\
&& m_{J/\psi}=3.0969\,{\rm GeV},  \;\;  \alpha(m_{J/\psi})=1/132.6,\nonumber \\
&& G_F=1.16638 \times 10^{-5}{\rm GeV}^{-2},
\end{eqnarray}
where $m_c$ stands for the pole mass of the charm quark, and its value is taken from Ref.\cite{Bodwin:2007fz}. The values of $m_{_Z}$, $m_{J/\psi}$, and $ G_F$ are taken from the Particle Data Group (PDG) \cite{ParticleDataGroup:2022pth}. The value of the running electromagnetic coupling $\alpha(m_{J/\psi})$ is taken from Ref.\cite{Bodwin:2007ga}. For the running strong coupling constant, we use the two-loop formula
\begin{eqnarray}
\alpha_s(\mu_r)=\frac{4\pi}{\beta_0\, {\rm ln}(\mu_r^2/\Lambda_{\rm QCD}^2)}\left[ 1-\frac{\beta_1 {\rm ln} \,{\rm ln}(\mu_r^2/\Lambda_{\rm QCD}^2)}{\beta_0^2\, {\rm ln}(\mu_r^2/\Lambda_{\rm QCD}^2)} \right],
\label{eq.alphas}
\end{eqnarray}
where $\beta_1=102-38 n_f/3$ is the two-loop coefficient of the QCD $\beta$ function. According to $\alpha_s(m_{_Z})=0.118$ \cite{ParticleDataGroup:2022pth}, we obtain $\Lambda^{n_f=5}_{\rm QCD}=0.226\, {\rm GeV}$ and $\Lambda^{n_f=4}_{\rm QCD}=0.328\, {\rm GeV}$. With the determined values of $\Lambda_{\rm QCD}$, the strong coupling constant at an arbitrary scale ($\mu_r \gg \Lambda_{\rm QCD}$) can be directly calculated through Eq.(\ref{eq.alphas}).

The values of $g_{J/\psi\gamma}$ and $f_{J/\psi}^\parallel$ can be extracted from the lepton decay width of $J/\psi$ through Eq.(\ref{eq.width-jpsi-lepton}) and Eq.(\ref{eq.EM-width-decay-constant}), respectively. Using the measured value $\Gamma(J/\psi \to e^{+}e^{-})=5.55 {\rm keV}$ \cite{ParticleDataGroup:2006fqo}, we obtain
\begin{eqnarray}
&& g_{J/\psi\gamma}=-0.832 \,{\rm GeV}^2, \\
&& f_{J/\psi}^\parallel=403\,{\rm MeV}.
\end{eqnarray}
The coupling constant $g_{J/\psi\gamma}$ has a relative minus sign compared to $f_{J/\psi}^\parallel$, which can be realized by comparing Eq.(\ref{eq.amp-gamma-jpsi}) and Eq.(\ref{eq.decay-constant-def}).

We take the values for the LDME $\langle \mathcal{O}_{1}\rangle_{J/\psi}=\frac{1}{3}\sum_{\lambda}\vert \langle J/\psi((\lambda)\vert \psi^\dagger {\bm \sigma}\cdot {\bm \epsilon} (\lambda)\chi \vert 0 \rangle \vert^2$ and the ratio of LDMEs $\langle {\bm q}^2\rangle_{J/\psi}$ from Ref.\cite{Bodwin:2007fz}, i.e.,
\begin{eqnarray}
&&\langle \mathcal{O}_{1}\rangle_{J/\psi}=0.440\,{\rm GeV}^{3}, \\
&&\langle {\bm q}^2\rangle_{J/\psi}=0.441\,{\rm GeV}^{2}.
\label{eq.ldmes}
\end{eqnarray}
The value of $\langle v^{2}\rangle_{J/\psi}$ can be derived through $\langle v^{2}\rangle_{J/\psi}=\langle \bm{q}^{2}\rangle_{J/\psi}/m_c^2$ directly.

\subsection{Basic results}

In this subsection, we present the numerical results for the decay width of the decay $Z\to J/\psi+J/\psi$. In order to have a glance on the magnitude of the photon fragmentation, non-fragmentation, and interference contributions to the decay width of $Z\to J/\psi+J/\psi$, we first present the numerical results with the input parameters taken as their central values. A detailed analysis of the uncertainty of the decay width will be presented in the next subsection.

\begin{table}[htb]
\caption{The decay width (in unit: $10^{-12}\,{\rm GeV}$) of the decay $Z \to J/\psi + J/\psi $. The photon fragmentation, non-fragmentation, interference, and total contributions are given explicitly.}
\begin{tabular}{c c c c c }
\hline\hline
  & Fragmentation  & Non-fragmentation &  Interference   & Total  \\
\hline
$\Gamma$  & $154.0 $   & $2.4 $  & $60.2$ & $216.6$  \\
\hline\hline
\end{tabular}
\label{tb.width}
\end{table}

Numerical results for the decay width are shown in Table \ref{tb.width}, where the photon fragmentation, non-fragmentation, interference, and total contributions are given explicitly. In the calculation, we have set the renormalization scale in the non-fragmentation amplitude as $\mu_r=m_{_Z}$.

From Table \ref{tb.width}, we can find that the largest contribution comes from the photon fragmentation, which accounts for $74.1\%$ of the total contribution. The interference of the photon fragmentation amplitude and the non-fragmentation amplitude accounts for $24.8\%$ of the total contribution, while the non-fragmentation contribution only accounts for $1.1\%$ of the total contribution. Although the photon fragmentation contribution is suppressed by the factor $(\alpha/\alpha_s)^2$ compared to the non-fragmentation contribution, the photon fragmentation contribution is two orders of magnitude lager than the non-fragmentation contribution. This is because the invariant mass of the photon propagator in the fragmentation diagrams is $m_{J/\psi}$, while the invariant mass of the gluon propagator in the LO non-fragmentation diagrams is $m_{_Z}/2$, i.e., the fragmentation contribution is enhanced by a factor of $(m_{_Z}/2m_{J/\psi})^4$ from the gauge boson propagator compared to the non-fragmentation contribution.

\subsection{Uncertainty analysis}

In this subsection, we present an estimate of the theoretical uncertainties for the decay width of $Z\to J/\psi+J/\psi$. The main uncertainty sources include the factorization/renormalization scales, the charm quark mass, the leptonic decay width $\Gamma(J/\psi \to e^{+}e^{-})$, the LDME $\langle \mathcal{O}_{1}\rangle_{J/\psi}$, and the ratio $\langle {\bm q}^2\rangle_{J/\psi}$. In the following estimation, when we discuss the uncertainty from one parameter, other parameters will be kept to be their central values.

There are several factorization/renormalization scales involved in the calculation: the initial factorization scale $\mu_0$ and the final factorization scale $\mu$ in the calculation of the photon fragmentation amplitude, as well as the renormalization scale $\mu_r$ in the calculation of the nonfragmentation amplitude. We estimate the uncertainties by varying these scales in the ranges $\mu_0 \in [1{\rm GeV}, 2m_c]$, $\mu \in [m_{_Z}/2,2m_{_Z}]$, and $\mu_r \in [m_{_Z}/2,2m_{_Z}]$. For the uncertainty caused by the charm quark mass, we estimate it by taking $m_c=1.4 \pm 0.2 \,{\rm GeV}$. It is noted that the values for the LDME $\langle \mathcal{O}_{1}\rangle_{J/\psi}$ and the ratio $\langle {\bm q}^2\rangle_{J/\psi}$ obtained in Ref.\cite{Bodwin:2007fz} depend on the value of the charm quark pole mass. Therefore, when considering the uncertainty caused by the charm quark mass, we not only consider the dependence of the SDCs on the charm quark mass, but also the dependence of the LDME $\langle \mathcal{O}_{1}\rangle_{J/\psi}$ and the ratio $\langle {\bm q}^2\rangle_{J/\psi}$ on the charm quark mass. The dependence of the LDME and the ratio $\langle {\bm q}^2\rangle_{J/\psi}$ on the charm quark mass has been given in the Ref.\cite{Bodwin:2007fz}. In our calculation, the parameters $g_{J/\psi\gamma}$ and $f_{J/\psi}^\parallel$ are extracted from the measured value of $\Gamma(J/\psi \to e^{+}e^{-})$, and the LDME $\langle \mathcal{O}_{1}\rangle_{J/\psi}$ and the ratio $\langle {\bm q}^2\rangle_{J/\psi}$ obtained in Ref.\cite{Bodwin:2007fz} also depend on the measured value of $\Gamma(J/\psi \to e^{+}e^{-})$. Thus, $\Gamma(J/\psi \to e^{+}e^{-})$ is an important uncertainty source. For the uncertainty caused by the measured value of $\Gamma(J/\psi \to e^{+}e^{-})$, we take the uncertainties of the leptonic width as $\Gamma(J/\psi \to e^{+}e^{-})=5.55\pm 0.14 \pm 0.02\, {\rm keV}$ \cite{ParticleDataGroup:2006fqo}. Since the uncertainties caused by the charm quark mass and the measured value of $\Gamma(J/\psi \to e^{+}e^{-})$ have already been considered, we will omit the uncertainties caused by charm quark mass and the measured value of $\Gamma(J/\psi \to e^{+}e^{-})$ when considering the uncertainties caused by the LDME $\langle \mathcal{O}_{1}\rangle_{J/\psi}$ and the ratio $\langle {\bm q}^2\rangle_{J/\psi}$. The uncertainties for the LDME $\langle \mathcal{O}_{1}\rangle_{J/\psi}$ and the ratio $\langle {\bm q}^2\rangle_{J/\psi}$ with omitting the uncertainties caused by the charm quark mass and the measured value of $\Gamma(J/\psi \to e^{+}e^{-})$ \cite{Bodwin:2007fz} are
\begin{eqnarray}
&&\langle \mathcal{O}_{1}\rangle_{J/\psi}=0.440^{+0.065}_{-0.054}\,{\rm GeV}^{3}, \\
&&\langle {\bm q}^2\rangle_{J/\psi}=0.441^{+0.139}_{-0.138}\,{\rm GeV}^{2}.
\end{eqnarray}
Then we obtain the uncertainties arising from these uncertainty sources
\begin{eqnarray}
&& \Gamma_{Z\to J/\psi+ J/\psi} \nonumber\\
&&=2.16^{+0.06 +0.32+0.12+0.10+0.07} _{-0.07 -0.05-0.10-0.08-0.06 }\times 10^{-10}\ {\rm GeV},
\end{eqnarray}
where the first uncertainty is caused by the factorization/renormalization scales \footnote{The uncertainties arising from the scales $\mu_0$, $\mu$, and $\mu_r$ have been added in quadrature.}, the second uncertainty is caused by the charm quark mass, the third uncertainty is caused by the measured value of $\Gamma(J/\psi \to e^{+}e^{-})$, the fourth uncertainty is cased by the LDME $\langle \mathcal{O}_{1}\rangle_{J/\psi}$ and the last uncertainty is caused by the ratio $\langle {\bm q}^2\rangle_{J/\psi}$. Adding these uncertainties in quadrature, we obtain the total theoretical
uncertainty for the decay width
\begin{eqnarray}
\Gamma_{Z\to J/\psi+ J/\psi}=2.16^{+0.37} _{-0.17}\times 10^{-10}\ {\rm GeV}.
\end{eqnarray}

Using the decay width obtained in the present paper and the total $Z$-boson decay width $\Gamma_{_Z}=2.4955\,{\rm GeV}$ from PDG \cite{ParticleDataGroup:2022pth}, we obtain the branching fraction for the decay $Z\to J/\psi+ J/\psi$
\begin{eqnarray}
{\rm Br}(Z \to J/\psi + J/\psi)=8.66 ^{+1.48} _{-0.69}\times 10^{-11}.
\label{eq.Br.psi}
\end{eqnarray}

\subsection{Comparison with other theoretical calculation and current experiment limit}

In this subsection, we present a comparison of our calculation with previous theoretical calculation and the experimental upper limit on the decay $Z\to J/\psi+ J/\psi$.

\begin{table}[h]
\caption{The branching fraction of the decay $Z\to J/\psi+J/\psi$. Our result and the result from Ref.\cite{Li:2023tzx} are theoretical predictions, while the limit in the last column is the current experimental upper limit given by Ref.\cite{CMS:2022fsq} on this branching fraction.}
\begin{center}
\begin{tabular}{c c c}
\hline\hline
~~Br(this work)~~& ~~${\rm Br}$(Ref.\cite{Li:2023tzx})~~ & ~~${\rm Br}$(Ref.\cite{CMS:2022fsq})~~ \\
\hline
$8.66 ^{+1.48} _{-0.69}\times 10^{-11}$  & $1.110^{+0.334+0.054}_{-0.241-0.001} \times 10^{-10} $ & $<1.4\times10^{-6}$ \\
\hline\hline
\end{tabular}
\end{center}
\label{tab.comparison}
\end{table}

The theoretical predictions and the current experimental upper limit on the branching fraction of the decay $Z\to J/\psi+ J/\psi$ are shown in Table \ref{tab.comparison}. From the table, we can see that our branching fraction for $Z\to J/\psi+ J/\psi$ is smaller than that in Ref.\cite{Li:2023tzx} by about $22\%$, which is about $-1.0 \sigma$ in the uncertainties of Ref.\cite{Li:2023tzx}.

In Ref.\cite{Li:2023tzx}, the photon fragmentation, non-fragmentation, and the interference contributions were calculated up to NLO in $\alpha_s$ based on the NRQCD factorization. It was found in Ref.\cite{Li:2023tzx} that the NLO QCD corrections are significant, e.g., the NLO QCD correction diminishes the photon fragmentation contribution by $20-40\%$ with different choices of the renormalization scale. In this work, the photon fragmentation amplitude is calculated with an improved approach. More explicitly, the photon fragmentation amplitude is written as the product of the amplitude of $Z\to J/\psi+\gamma^*$ and the amplitude of $\gamma^* \to J/\psi$. The amplitude of $\gamma^* \to J/\psi$ is extracted from the measured value of $\Gamma(J/\psi \to e^{+}e^{-})$, and the amplitude of $Z\to J/\psi+\gamma^*$ is calculated through the light-cone approach where the large logarithms of $m_{_Z}^2/m_c^2$ are resummed through the evolution of the $J/\psi$ LCDA. Using this improved approach, the higher-order QCD and relativistic corrections in the amplitude of $\gamma^* \to J/\psi$ and the large logarithms of $m_{_Z}^2/m_c^2$ in the amplitude of $Z\to J/\psi+\gamma^*$ are resummed in our calculation. Moreover, in our calculation of the non-fragmentation amplitude, besides the NLO QCD corrections, the NLO relativistic corrections are also included.

From Table \ref{tab.comparison}, we can see that our branching fraction for $Z\to J/\psi+ J/\psi$ is compatible with the current experimental upper limit established by the CMS Collaboration.

\section{summary}
\label{sec4}

In this paper, we have presented a calculation for the decay rate of $Z\to J/\psi+ J/\psi$. The decay process $Z\to J/\psi+ J/\psi$ is dominated by the single photon fragmentation mechanism, i.e., the transition $Z \to J/\psi+\gamma^*$ followed by the fragmentation $\gamma^* \to J/\psi$. The amplitude of $\gamma^* \to J/\psi$ has already been calculated up to ${\cal O}(\alpha_s^3)$ under the NRQCD factorization in Refs.\cite{Marquard:2014pea,Egner:2022jot}. It seems that the perturbative expansion of the amplitude of $\gamma^* \to J/\psi$ does not converge in those calculations. Moreover, the higher-order corrections of the amplitude of $Z \to J/\psi+\gamma^*$ involve the large logarithms of $m_{_Z}^2/m_c^2$. Considering these problems, we have presented an improved calculation for the decay rate of $Z\to J/\psi+ J/\psi$ compared to the fixed-order calculation. In our calculation, the amplitude for the photon fragmentation is expressed as the product of the amplitude of $Z \to J/\psi+\gamma^*$ and the amplitude of $\gamma^* \to J/\psi$. Then, the amplitude of $\gamma^* \to J/\psi$ is extracted from the measured value of $\Gamma(J/\psi \to e^{+}e^{-})$, and the amplitude of $Z \to J/\psi+\gamma^*$ is calculated by the light-cone approach where the large logarithms of $m_{_Z}^2/m_c^2$ are resummed through solving the evolution equation of the $J/\psi$ LCDA. The advantage of our calculation is that the higher-order QCD and relativistic corrections in the amplitude of $\gamma^* \to J/\psi$ and the large logarithms of $m_{_Z}^2/m_c^2$ in the amplitude of $Z \to J/\psi+\gamma^*$ are resummed. For the non-fragmentation amplitude, we adopt the fixed-order approach under the NRQCD factorization. In our calculation of the non-fragmentation amplitude, the NLO QCD and NLO relativistic corrections are included simultaneously.

The obtained branching fraction for $Z\to J/\psi+ J/\psi$ is $8.66 ^{+1.48} _{-0.69}\times 10^{-11}$, which is compatible with the current experimental upper limit ${\rm Br}(Z \to J/\psi + J/\psi)=1.4\times 10^{-6}$ given by the CMS collaboration \cite{CMS:2022fsq}. In recent years, several high-luminosity $e^+e^-$ colliders, such as ILC, CEPC, FCC-ee, and Super $Z$ factory are proposed. A large number of $Z$ boson events are expected to be generated at these future $e^+e^-$ colliders. For instance, about $5\times 10^{12}$ $Z$ bosons can be generated at the FCC-ee \cite{FCC:2018evy}. According to the branching fraction obtained in this work, about $400$ $Z\to J/\psi+ J/\psi$ events are expected to be generated at the FCC-ee. One may expect that this decay can be studied at these future $e^+e^-$ colliders.

\hspace{2cm}

\noindent {\bf Acknowledgments:} We thank Hai-Bing Fu for helpful discussions on the evolution of the light-cone distribution amplitude of $J/\psi$. This work was supported in part by the Natural Science Foundation of China under Grants No. 12005028, No. 12175025, No. 12275036 and No. 12347101, by the Fundamental Research Funds for the Central Universities under Grant No. 2020CQJQY-Z003, by the Chongqing Natural Science Foundation under Grants No. CSTB2022NSCQ-MSX0415 and  No. cstc2021jcyj-msxmX0681, and by the Chongqing Graduate Research and Innovation Foundation under Grant No. ydstd1912.

\hspace{2cm}


\begin{thebibliography}{1}

\bibitem{CMS:2019wch}
A.~M.~Sirunyan \textit{et al.} [CMS],
Search for Higgs and $Z$ boson decays to J/\ensuremath{\psi} or Y pairs in the four-muon final state in proton-proton collisions at s=13TeV,
Phys. Lett. B \textbf{797}, 134811 (2019).

\bibitem{CMS:2022fsq}
A.~Tumasyan \textit{et al.} [CMS],
Search for Higgs boson decays into Z and J/\ensuremath{\psi} and for Higgs and $Z$ boson decays into J/\ensuremath{\psi} or Y pairs in pp collisions at s=13~TeV,
Phys. Lett. B \textbf{842}, 137534 (2023).

\bibitem{dEnterria:2023wjq}
D.~d'Enterria and V.~Le,
Rare and exclusive few-body decays of the Higgs, Z, W bosons, and the top quark,
arXiv:2312.11211 [hep-ph].

\bibitem{ILC:2013jhg}
H.~Baer \textit{et al.} [ILC],
The International Linear Collider Technical Design Report - Volume 2: Physics,
arXiv:1306.6352 [hep-ph].

\bibitem{CEPCStudyGroup:2018ghi}
J.~B.~Guimar\~aes da Costa \textit{et al.} [CEPC Study Group],
CEPC Conceptual Design Report: Volume 2 - Physics \& Detector,
arXiv:1811.10545 [hep-ex].

\bibitem{FCC:2018evy}
A.~Abada \textit{et al.} [FCC],
FCC-ee: The Lepton Collider: Future Circular Collider Conceptual Design Report Volume 2,
Eur. Phys. J. ST \textbf{228}, 261-623 (2019).

\bibitem{zfactory}
J. P. Ma and Z. X. Zhang (The super Z-factory group), Preface,
Sci. China: Phys., Mech. Astron. {\bf 53}, 1947 (2010).

\bibitem{Bergstrom:1990bu}
L.~Bergstrom and R.~W.~Robinett,
ON THE RARE DECAYS $Z \to V V$ AND $Z \to V P$,
Phys. Rev. D \textbf{41}, 3513 (1990).

\bibitem{Likhoded:2017jmx}
A.~K.~Likhoded and A.~V.~Luchinsky,
Double Charmonia Production in Exclusive $Z$ Boson Decays,
Mod. Phys. Lett. A \textbf{33}, 1850078 (2018).

\bibitem{Gao:2022mwa}
D.~N.~Gao and X.~Gong,
Note on rare Z-boson decays to double heavy quarkonia,
Chin. Phys. C \textbf{47}, 043106 (2023).

\bibitem{Li:2023tzx}
C.~Li, Z.~Sun and G.~Y.~Zhang,
Analysis of double-J/\ensuremath{\psi} production in Z decay at next-to-leading-order QCD accuracy,
JHEP \textbf{10}, 120 (2023).

\bibitem{Luo:2022ugd}
X.~Luo, H.~B.~Fu, H.~J.~Tian and C.~Li,
Next-to-leading-order QCD correction to the exclusive double charmonium production via $Z$ decays,
arXiv:2209.08802 [hep-ph].

\bibitem{Lepage:1980fj}
G.~P.~Lepage and S.~J.~Brodsky,
Exclusive Processes in Perturbative Quantum Chromodynamics,
Phys. Rev. D \textbf{22}, 2157 (1980).

\bibitem{Chernyak:1983ej}
V.~L.~Chernyak and A.~R.~Zhitnitsky,
Asymptotic Behavior of Exclusive Processes in QCD,
Phys. Rept. \textbf{112}, 173 (1984).

\bibitem{Bodwin:1994jh}
G.~T.~Bodwin, E.~Braaten and G.~P.~Lepage,
Rigorous QCD analysis of inclusive annihilation and production of heavy quarkonium,
Phys. Rev. D \textbf{51}, 1125-1171 (1995).

\bibitem{Bodwin:2006yd}
G.~T.~Bodwin, E.~Braaten, J.~Lee and C.~Yu,
Exclusive two-vector-meson production from e+ e- annihilation,
Phys. Rev. D \textbf{74}, 074014 (2006).

\bibitem{Bodwin:2013gca}
G.~T.~Bodwin, F.~Petriello, S.~Stoynev and M.~Velasco,
Higgs boson decays to quarkonia and the $H\bar{c}c$  coupling,
Phys. Rev. D \textbf{88}, 053003 (2013).

\bibitem{Guberina:1980dc}
B.~Guberina, J.~H.~Kuhn, R.~D.~Peccei and R.~Ruckl,
Rare Decays of the Z0,
Nucl. Phys. B \textbf{174}, 317 (1980).

\bibitem{Luchinsky:2017jab}
A.~V.~Luchinsky,
Leading order NRQCD and Light-Cone Analysis of Exclusive Charmonia Production in Radiative $Z$-boson Decays,
arXiv:1706.04091.

\bibitem{Wang:2013ywc}
X.~P.~Wang and D.~Yang,
The leading twist light-cone distribution amplitudes for the S-wave and P-wave quarkonia and their applications in single quarkonium exclusive productions,
JHEP \textbf{06}, 121 (2014).

\bibitem{Bodwin:2017pzj}
G.~T.~Bodwin, H.~S.~Chung, J.~H.~Ee and J.~Lee,
$Z$-boson decays to a vector quarkonium plus a photon,
Phys. Rev. D \textbf{97}, 016009 (2018).

\bibitem{Sang:2023hjl}
W.~L.~Sang, D.~S.~Yang and Y.~D.~Zhang,
$Z$-boson radiative decays to an S-wave quarkonium at NNLO and NLL accuracy,
Phys. Rev. D \textbf{108}, 014021 (2023).

\bibitem{Wang:2023ssg}
G.~Y.~Wang, X.~C.~Zheng, X.~G.~Wu and G.~Z.~Xu,
$Z$-boson decays into $S$-wave quarkonium plus a photon up to ${\cal O}(\alpha_{s} v^2)$ corrections,
Phys. Rev. D \textbf{109}, 074004 (2024).

\bibitem{Ball:1996tb}
P.~Ball and V.~M.~Braun,
The Rho meson light cone distribution amplitudes of leading twist revisited,
Phys. Rev. D \textbf{54}, 2182-2193 (1996).

\bibitem{Ball:1998sk}
P.~Ball, V.~M.~Braun, Y.~Koike and K.~Tanaka,
Higher twist distribution amplitudes of vector mesons in QCD: Formalism and twist - three distributions,
Nucl. Phys. B \textbf{529}, 323-382 (1998).

\bibitem{Ma:2006hc}
J.~P.~Ma and Z.~G.~Si,
NRQCD Factorization for Twist-2 Light-Cone Wave-Functions of Charmonia,
Phys. Lett. B \textbf{647}, 419-426 (2007).

\bibitem{Wang:2017bgv}
W.~Wang, J.~Xu, D.~Yang and S.~Zhao,
Relativistic corrections to light-cone distribution amplitudes of S-wave B$_{c}$ mesons and heavy quarkonia,
JHEP \textbf{12}, 012 (2017).

\bibitem{Efremov:1979qk}
A.~V.~Efremov and A.~V.~Radyushkin,
Factorization and Asymptotical Behavior of Pion Form-Factor in QCD,
Phys. Lett. B \textbf{94}, 245-250 (1980).

\bibitem{Efremov:1978rn}
A.~V.~Efremov and A.~V.~Radyushkin,
Asymptotical Behavior of Pion Electromagnetic Form-Factor in QCD,
Theor. Math. Phys. \textbf{42}, 97-110 (1980).

\bibitem{Dittes:1983dy}
F.~M.~Dittes and A.~V.~Radyushkin,
Two loop contribution to the evolution of the pion wave function,
Phys. Lett. B \textbf{134}, 359-362 (1984).

\bibitem{Katz:1984gf}
G.~R.~Katz,
Two Loop Feynman Gauge Calculation of the Meson Nonsinglet Evolution Potential,
Phys. Rev. D \textbf{31}, 652 (1985).

\bibitem{Mikhailov:1984ii}
S.~V.~Mikhailov and A.~V.~Radyushkin,
Evolution Kernels in {QCD}: Two Loop Calculation in Feynman Gauge,
Nucl. Phys. B \textbf{254}, 89-126 (1985).

\bibitem{Agaev:2010aq}
S.~S.~Agaev, V.~M.~Braun, N.~Offen and F.~A.~Porkert,
Light Cone Sum Rules for the pi0-gamma*-gamma Form Factor Revisited,
Phys. Rev. D \textbf{83}, 054020 (2011).


\bibitem{Bodwin:2016edd}
G.~T.~Bodwin, H.~S.~Chung, J.~H.~Ee and J.~Lee,
New approach to the resummation of logarithms in Higgs-boson decays to a vector quarkonium plus a photon,
Phys. Rev. D \textbf{95}, 054018 (2017).

\bibitem{Bodwin:2010fi}
G.~T.~Bodwin, X.~Garcia Tormo, i and J.~Lee,
Factorization in exclusive quarkonium production,
Phys. Rev. D \textbf{81}, 114014 (2010).

\bibitem{Beneke:1997zp}
M.~Beneke and V.~A.~Smirnov,
Asymptotic expansion of Feynman integrals near threshold,
Nucl. Phys. B \textbf{522}, 321 (1998).

\bibitem{Larin:1993tq}
S.~A.~Larin,
The Renormalization of the axial anomaly in dimensional regularization,
Phys. Lett. B \textbf{303}, 113-118 (1993).

\bibitem{feynarts}
T. Hahn,
Generating Feynman diagrams and amplitudes with FeynArts 3,
Comput. Phys. Commun {\bf 140}, 418 (2001).

\bibitem{feyncalc1}
R. Mertig, M. Bohm and A. Denner,
FeynCalc - Computer-algebraic calculation of Feynman amplitudes,
Comput. Phys. Commun {\bf 64}, 345 (1991).

\bibitem{feyncalc2}
V. Shtabovenko, R. Mertig and F. Orellana,
New Developments in FeynCalc 9.0,
Comput. Phys. Commun {\bf 207}, 432 (2016).

\bibitem{apart}
F. Feng,
\$Apart: A Generalized Mathematica Apart Function,
Comput. Phys. Commun {\bf 183}, 2158 (2012).

\bibitem{fire}
A.V. Smirnov,
Algorithm FIRE - Feynman Integral REduction,
J. High Energy Phys. {\bf 0810}, 107 (2008).

\bibitem{looptools}
T. Hahn and M. Perez-Victoria,
Automatized one loop calculations in four-dimensions and D-dimensions,
Comput. Phys. Commun {\bf 118}, 153 (1999).

\bibitem{Bodwin:2007fz}
G.~T.~Bodwin, H.~S.~Chung, D.~Kang, J.~Lee and C.~Yu,
Improved determination of color-singlet nonrelativistic QCD matrix elements for S-wave charmonium,
Phys. Rev. D \textbf{77}, 094017 (2008).

\bibitem{ParticleDataGroup:2022pth}
R.~L.~Workman \textit{et al.} [Particle Data Group],
Review of Particle Physics,
PTEP \textbf{2022}, 083C01 (2022).

\bibitem{Bodwin:2007ga}
G.~T.~Bodwin, J.~Lee and C.~Yu,
Resummation of Relativistic Corrections to $e+ e- \to J/psi + eta(c)$,
Phys. Rev. D \textbf{77}, 094018 (2008).

\bibitem{ParticleDataGroup:2006fqo}
W.~M.~Yao \textit{et al.} [Particle Data Group],
Review of Particle Physics,
J. Phys. G \textbf{33}, 1-1232 (2006).

\bibitem{Marquard:2014pea}
P.~Marquard, J.~H.~Piclum, D.~Seidel and M.~Steinhauser,
Three-loop matching of the vector current,
Phys. Rev. D \textbf{89}, 034027 (2014).

\bibitem{Egner:2022jot}
M.~Egner, M.~Fael, F.~Lange, K.~Sch\"onwald and M.~Steinhauser,
Three-loop nonsinglet matching coefficients for heavy quark currents,
Phys. Rev. D \textbf{105}, 114007 (2022).

\end{thebibliography}
\end{document}